\documentclass{Interspeech}

\interspeechcameraready

\title{Decoding Speaker-Normalized Pitch from EEG for Mandarin Perception}

\author[affiliation={1,2}]{Jiaxin}{Chen}
\author[affiliation={1,2}]{Yiming}{Wang}
\author[affiliation={2}]{Ziyu}{Zhang}
\author[affiliation={2}]{Jiayang}{Han}
\author[affiliation={1,2}]{Yin-Long}{Liu}
\author[affiliation={1,2}]{Rui}{Feng}
\author[affiliation={2}]{Xiuyuan}{Liang}
\author[affiliation={1,2}]{Zhen-Hua}{Ling}
\author[affiliation={1,2}]{Jiahong}{Yuan}

\affiliation{National Engineering Research Center of Speech and Language Information Processing}{University of Science and Technology of China}{Hefei, P. R. China}
\affiliation{Interdisciplinary Research Center for Linguistic Sciences}{University of Science and Technology of China}{Hefei, P. R. China}
\email{jxchen4759@mail.ustc.edu.cn, \{zhling, jiahongyuan\}@ustc.edu.cn}


\usepackage{comment}
\usepackage{tabularx}
\usepackage{hyperref} 

\begin{document}

\maketitle

\begin{abstract}
    
    The same speech content produced by different speakers exhibits significant differences in pitch contour, yet listeners' semantic perception remains unaffected. This phenomenon may stem from the brain's perception of pitch contours being independent of individual speakers' pitch ranges. In this work, we recorded electroencephalogram (EEG) while participants listened to Mandarin monosyllables with varying tones, phonemes, and speakers. The CE-ViViT model is proposed to decode raw or speaker-normalized pitch contours directly from EEG. Experimental results demonstrate that the proposed model can decode pitch contours with modest errors, achieving performance comparable to state-of-the-art EEG regression methods. Moreover, speaker-normalized pitch contours were decoded more accurately, supporting the neural encoding of relative pitch.
    
\end{abstract}


\section{Introduction}
\vspace{0.1cm}

The fundamental frequency (F0), often referred to as pitch, is an essential and critical acoustic feature in speech signals, representing the lowest frequency component of a sound.
Humans precisely control the pitch of their voices to encode linguistic meaning \cite{honorof2005perception}.
In non-tonal languages like English, pitch variations convey intonational information, including emotions, attitudes, questions and so on.
In contrast, tonal languages such as Mandarin, pitch changes create distinct tones that carry different semantic meanings \cite{chao1930system}.
Thus, pitch as a critical acoustic correlate for Mandarin tones, as the systematic variations in pitch contours enable the perceptual distinction among different tonal categories \cite{yuan2023improved}.
For different speakers, the range and variation of pitch contours may differ significantly, leading to considerable variability in the pronunciation of the same syllable \cite{honorof2005perception,henton1995pitch}.
However, the mechanisms underlying the auditory perception of pitch remain poorly undared. 
Investigating how the brain interprets semantic meaning from speech can help guide precise preservation of language functions in neurosurgery \cite{chung2023decoding}.

Over the past decade, invasive brain-computer interface (BCI) technologies have begun to be utilized for precise mapping of brain circuits, 
enabling the study of speech production \cite{si2017cooperative,tang2017intonational,li2021human} and perception mechanisms \cite{liu2023decoding,shigemi2023synthesizing}, such as through electrocorticography (ECoG).
For instance, a study \cite{si2017cooperative} revealed categorical neural responses to lexical tones within a distributed network, encompassing not only auditory regions in the temporal cortex but also motor areas in the frontal cortex.
Liu et al. \cite{liu2023decoding} designed a multistream neural network that independently decodes lexical tones and base syllables from intracranial recordings to synthesize tonal language speech, using parallel streams inspired by neuroscience findings.

However, invasive methods require complex brain surgeries, which are costly and difficult to replicate or scale. 
As a result, some teams have focused on decoding speech from non-invasive recordings of brain activity like electroencephalography (EEG) \cite{defossez2023decoding,ni2024dbpnet,accou2021modeling,postolache2024naturalistic}.
Wang et al. \cite{wang2023cross} utilized an end-to-end convolutional neural network (CNN) model to perform classification tasks for Mandarin speech directly from raw EEG signals, reporting average classification accuracies of 63.1\% for four vowels and 51.7\% for four tones.
Furthermore, several studies have dedicated efforts to regression tasks, specifically targeting the direct decoding of speech stimulus envelopes \cite{accou2023decoding, piao2023happyquokka} or mel-spectrograms \cite{xu2024convconcatnet} from EEG data.
Despite existing challenges associated with these non-invasive methods \cite{hamalainen1993magnetoencephalography, king2018encoding}, the simplicity and cost-effectiveness of EEG have fostered its widespread adoption and steady progress in the field.

Recent study \cite{tang2017intonational} have investigated how the auditory cortex processes intonational pitch in speech, designing and synthesizing controlled spoken sentences with independent variations in intonation contour, phonetic content, and speaker identity, while using ECoG to directly record neural activity on the brain surface during auditory perception.
Research indicates that the neural encoding of pitch contours prioritizes speaker-normalized relative pitch over absolute pitch, highlighting the brain's focus on relative pitch changes rather than specific values.
However, it remains unclear whether speaker normalization for pitch occurs within the context of sentences or originates from the phonemes themselves. 
Additionally, ECoG is limited to capturing data from only a small subset of brain regions, with relatively sparse data availability.

\begin{figure*}[th]
  \centering
  \includegraphics[width=0.99\textwidth]{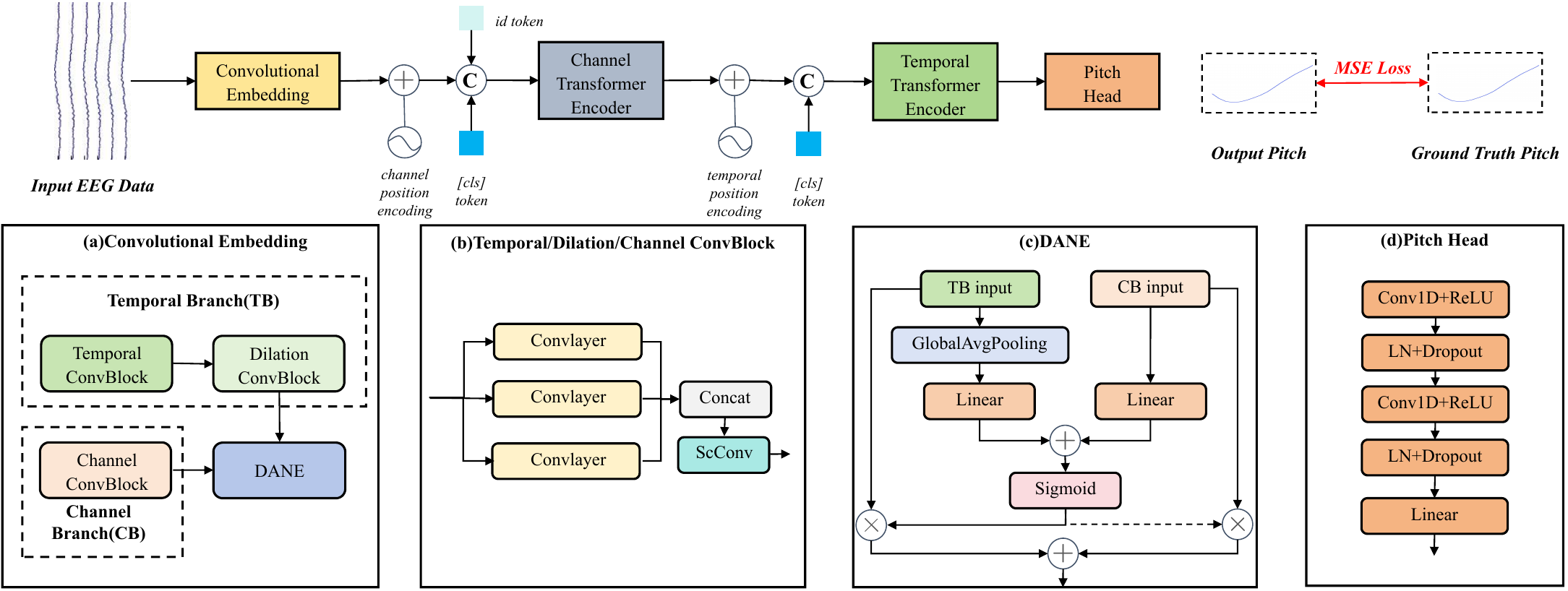} 
  \caption{The overall architecture for CE-ViViT model: (a) The dual parallel design of the temporal and channel domains in Convolutional Embedding (CE). (b) The details of the Temporal, Dilation, and Channel ConvBlocks. (c) The feature fusion method of DANE. (d) Illustration of the Pitch Head.}
  \label{fig:speech_production}
\end{figure*}

In this work, we aim to validate this theory by recording a more extensive auditory EEG dataset of monosyllabic speech, 
encompassing a wider variety of phonetic and tonal variations. 
To this end, we developed a nonlinear model, termed CE-ViViT, 
which processes electroencephalogram (EEG) signal segments as input. 
The model employs a multi-scale convolutional module for feature extraction, integrates features using ScConv \cite{li2023scconv} and DANE \cite{gao2022doubly}, decodes auditory attention through ViViT \cite{arnab2021vivit}, and ultimately maps pitch contours. 
We reconstructed both raw and speaker-normalized pitch contours from monosyllabic Mandarin speech produced by a single speaker and multiple speakers, respectively. 
The proposed model achieves performance on par with existing state-of-the-art auditory EEG decoding systems. 
Experimental results indicate that all models more effectively decode speaker-normalized pitch from EEG when multiple speakers are involved, while no significant difference is observed between speaker-normalized and raw pitch for a single speaker. 
This is consistent with the findings reported in \cite{tang2017intonational}.

\section{Proposed Method}
\vspace{0.1cm}

\subsection{Overview}
The overall structure of the proposed CE-ViViT is illustrated in Figure \ref{fig:speech_production}, which consists of three main blocks: Convolutional Embedding, Feature Encoder, and Pitch Head.
Raw EEG segments \( X \in \mathbb{R}^{C \times T} \) serve directly as input, where \( C \) represents the number of EEG channels and \( T \) denotes the number of sampling points in the signal.

The Convolutional Embedding module, positioned at the beginning of CE-ViViT, consists of parallel multi-scale convolutional layers stacked together, designed to extract local spatiotemporal features. 
It transforms the input \( X \) into an embedding \( V \in \mathbb{R}^{C' \times T' \times D} \), where \( D \) represents the dimensionality of the extracted spatiotemporal representations, and \( T' \) and \( C' \) denote the reduced temporal and channel dimensions, respectively.
It is followed by the Feature Encoder, which comprises a series of Transformer blocks designed to learn global context information. 
Here, temporal and channel domain information are fused, ultimately producing the feature representation \( O \in \mathbb{R}^{T' \times D} \).

Finally, the output latent vector \( O \) is aggregated by the Pitch Head into the pitch contour \( \hat{F_0} \in \mathbb{R}^{t} \), where \( t \) represents the number of time points in the pitch contour.
As illustrated in Figure \ref{fig:speech_production}(d), the Pitch Head consists of a 2-layer 1D convolutional network with ReLU activation, followed by a layer normalization and dropout layer after each network, as well as an additional linear layer that projects the hidden states into the output sequence.
The system's loss function is the mean squared error (MSE) loss between the reconstructed pitch contours \(\hat{F}_0\) and the ground-truth pitch contours \(F_0\), as shown in the following formula:
\begin{align}
    \mathcal{L}_{\text{pitch}} = \frac{1}{N} \sum_{i=1}^{N} (\hat{F}_0 - F_0)^2
\end{align}
where \(N\) is the total number of samples in the batch.

\subsection{Convolutional Embedding}
As showed in Figure \ref{fig:speech_production}(a), the Convolutional Embedding architecture comprises two parallel 2D convolutions designed to extract temporal and channel domain features, respectively, which are subsequently fused to produce the embedding \( V \). 
The temporal branch adopts a two-level cascaded structure, integrating a Temporal ConvBlock followed by a Dilation ConvBlock. 
Both blocks share an identical design, featuring multiple layers of parallel 2D convolutions with varying kernel sizes or dilation factors, and a single ScConv layer is employed to refine the resulting feature maps. 
In contrast, due to the relatively low resolution of the EEG channel domain, the channel branch consists solely of a Channel ConvBlock. 
This block utilizes multiple parallel convolutional filters with varying kernel sizes to extract parameters, followed by an ScConv layer to refine the channel domain feature maps. 
Finally, this work incorporates DANE \cite{gao2022doubly} from the Doubly Fused ViT (DFvT), applying separate attention mechanisms to both temporal and channel outputs to generate the embedding \( V \). 
The detailed architecture of DANE is illustrated in Figure \ref{fig:speech_production}(c).

The Temporal ConvBlock, Dilation ConvBlock, and Channel ConvBlock share the same structure, each using parallel convolutional filters with variations in temporal kernel sizes, channel dimensions, or dilation factors, respectively. The detailed architecture is shown in Figure \ref{fig:speech_production}(b).
ScConv \cite{li2023scconv}, an efficient module for visual tasks, reduces spatial and channel redundancies in CNNs. 
It includes an SRU for spatial reduction and a CRU for channel reduction. 


\subsection{Feature Encoder}
The embedding vector \( V \), with both temporal and channel dimensions, is encoded using the Video Vision Transformer (ViViT) \cite{arnab2021vivit}, as shown in Figure \ref{fig:speech_production}. 
Specifically, this work adopts ViViT Model 2, a two-stage encoder comprising two distinct Transformer Encoders.
In the first stage, the channel encoder models features from the same temporal index, with the output on the [cls] token representing the optimal channel combination derived through multi-layer attention. 
The temporal encoder, which receives input containing only the temporal and representation dimensions, learns global temporal domain information. 
The final output integrates representations from both temporal and spatial domains.

Each encoder layer includes layer normalization, multi-head self-attention, feed-forward layers, and residual connections, following the standard Transformer architecture \cite{vaswani2017attention}. 
Before inputting data into ViViT's channel and temporal encoders, channel and temporal positional encodings are added, using a sinusoidal and cosine scheme. 
Additionally, the identity vector \( s \) is concatenated along the channel dimension before feeding into the channel encoder.

  

%
%

\section{Experiments}
\vspace{0.1cm}

\subsection{Dataset and preprocessing}
The stimuli comprised two monosyllabic sessions, each containing 480 Mandarin monosyllabic words selected from the Tone Perfect corpus\footnote{https://tone.lib.msu.edu} \cite{TonePerfect}.  The sessions are described as follows:
\begin{itemize}
\item \textbf{Session 1:} Included 480 unique monosyllabic tokens produced by a single native Mandarin male speaker, covering all four lexical tones (120 tokens per tone).
\item \textbf{Session 2:} Consisted of 480 monosyllabic sounds produced by six native Mandarin speakers (three female and three male), with each speaker producing 20 monosyllabic sounds across all four tones (80 tokens per speaker).
\end{itemize}

The selected words included pseudowords (combinations of vowels, consonants, and tones that do not exist in Mandarin). All monosyllabic sounds were shorter than 1 second.
Here, we extracted the pitch contour using the PyWorld Vocoder \footnote{https://github.com/JeremyCCHsu/Python-Wrapper-for-World-Vocoder} and applied z-score normalization to all samples for each speaker.

The participants in the experiment were 29 healthy college students, including 19 males and 10 females. 
The participants’ ages ranged from 19 to 29 years, and all of them had no auditory or neurological disorders, normal hearing, and used Mandarin in daily communication.

Participants sat in a comfortable sofa in an electrically shielded soundproof room, during whole-head EEG recording. 
The stimuli were binaurally presented through headphones (Sennheiser HD 25) at an intensity of ~70 dB sound pressure level.
Each monosyllabic stimulus was presented once with a fixed inter-stimulus interval of 1.2 seconds, and the sequence of stimuli was kept identical for all participants to maintain consistency across the study.

Whole-head EEG signals were recorded using a SynAmps RT amplifier (NeuroScan, Charlotte, NC, USA) with a cap carrying 64 Ag/AgCl electrodes placed on the scalp at specific locations according to the extended international 10-20 system. 
Data were recorded at a sampling rate of 500 Hz. 
The reference electrode was attached to the tip of the nose, and electrode AFz was used as the ground electrode during recording. 
To control for eye-movement artifacts, horizontal and vertical eye movements were recorded using two bipolar electrooculography (EOG) electrodes. 
All electrode impedances were maintained below \(5\,\text{k}\Omega\). 

All EEG data were preprocessed via EEGLAB \cite{delorme2004eeglab}, including global average re-referencing and FIR bandpass filtering (0.5–40 Hz). 
Artifacts (e.g., ECG, EOG, EMG) were removed using independent component analysis (ICA), following established EEG protocols \cite{krishna2020speech}. 
For temporal alignment with monosyllabic stimuli, EEG segments exceeding speech duration were zero-padded and trimmed to 1s to cover core pronunciation phases. Data were standardized into matrices of dimensions (500, 60) and normalized for model input consistency.

\subsection{Experimental Setup}

Given our objective to investigate the brain’s perception of speaker-normalized pitch contours, the main experiments were conducted on session 2, which involved six speakers, to compare the model’s decoding performance on raw and speaker-normalized pitch contours. 
Session 1, which included data from only one speaker, was primarily utilized for assessing model performance and conducting ablation experiments.

The EEG data from 2 male and 1 female participants were designated as the target subjects, while the data from the remaining 26 participants were used in the training process. The dataset was partitioned into a training set and a validation set, with 80\% of the data allocated for training and the remaining 20\% for validation.
During training, the Adam optimizer was employed with a learning rate of 5e-4, combined with a CosineAnnealingLR scheduler set to a cycle of 20 and an eta\_min of 5e-6. 
To mitigate overfitting, a dropout rate of 0.2 was applied to each layer. The model was trained for 150 epochs with a batch size of 32, and the best-performing model on the validation set was selected.


\subsection{Evaluation Metrics}
This paper employs three evaluation metrics: Mean Absolute Error (MAE), Relative Mean Absolute Error (RMAE), and Gross Pitch Error (GPE) \cite{cheng1975comparative}. 

The MAE measures the average of absolute differences between predicted and true values:
\begin{align}
MAE = \frac{1}{N} \sum_{t=1}^{N} |\hat{F}_0 - F_0|
\end{align}
where \( N \) is the total number of voiced frames, \( \hat{F}_0 \) is the predicted pitch contours, and \( F_0 \) is the ground-truth pitch contours.

The RMAE calculates the average of relative errors between predicted and true values:
\begin{align}
    RMAE = \frac{1}{N} \sum_{t=1}^{N} \frac{|\hat{F}_0 - F_0|}{F_0}
\end{align}

The Gross Pitch Error (GPE) is defined as the percentage of voiced frames where the relative error in the estimated pitch surpasses 20\% of the reference true value. The formula is as follows:
\begin{align}
    GPE_{0.2}=\frac{N_{GPE}}{N_v} \times 100\%
\end{align}

where $N_{GPE}$ is the number of voiced frames with an estimated pitch relative error exceeding 20\% of the ground truth, and $N_v$ is the total number of voiced frames.

In subsequent experiments, we observe that these three metrics generally yield consistent results. The discussions will primarily focus on the MAE metric.

\section{Results and Analysis}
\vspace{0.1cm}

\subsection{Main Results}
\begin{figure}[th]
  \centering
  \includegraphics[width=0.45\textwidth]{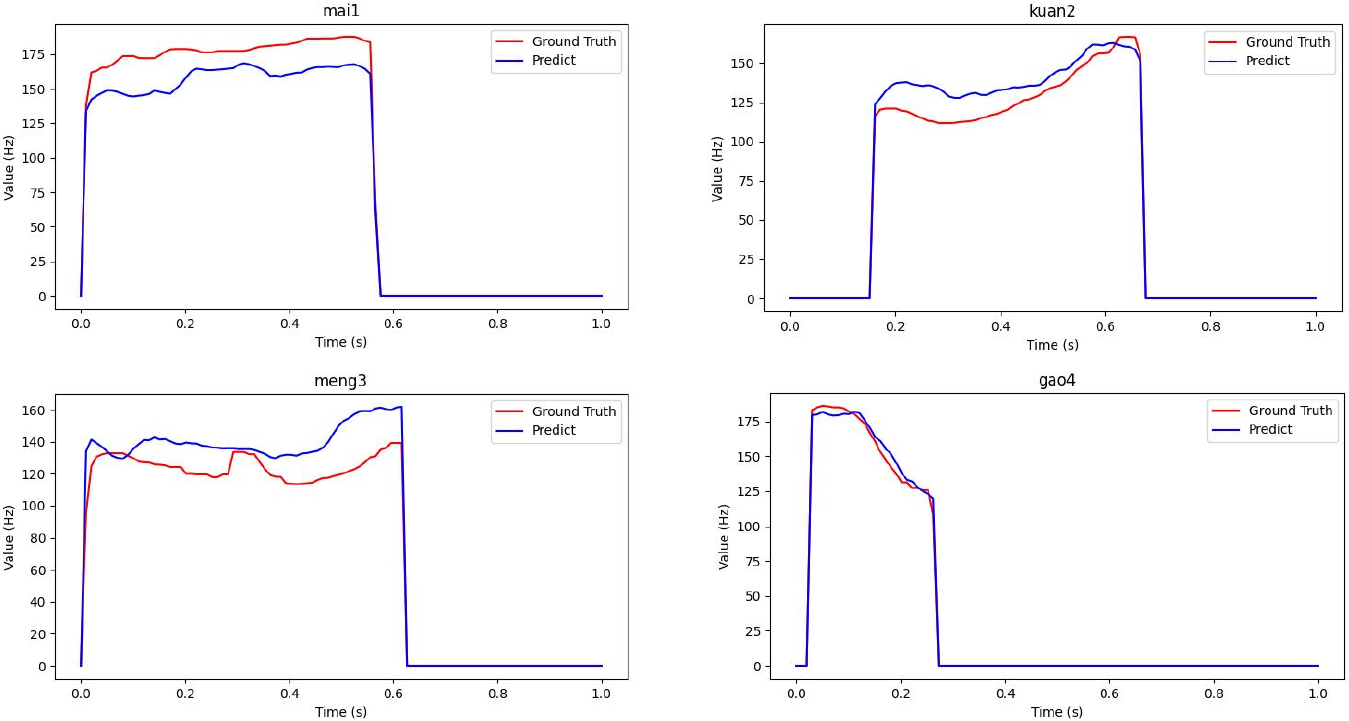} 
  \caption{Some examples of EEG-guided pitch reconstruction on session 1.}
  \label{fig:sample}
\end{figure}

As shown in Figure \ref{fig:sample}, these are several examples of the Mandarin pitch contours decoded from EEG by our proposed model on session 1.
For performance comparison, we consider VLAAI \cite{accou2023decoding}, WaveNet \cite{van2023decoding}, and HappyQuokka \cite{piao2023happyquokka} as contrastive systems, which are representative SOTA EEG-based speech regression models.

To address the significant scale differences between speaker-normalized pitch contours and raw pitch contours, 
which complicate direct comparisons of evaluation metrics, speaker-normalized pitch contours are rescaled to the original raw pitch contours' scale using the speaker's mean and variance during metric calculation. 
However, the training objective remains focused on speaker-normalized pitch contours.

First, Table \ref{tab:session1} reveals no significant performance difference between speaker-normalized and raw pitch contours in the single-speaker dataset, ruling out the possibility that performance gains stem solely from reduced dynamic range due to normalization. 
Second, as demonstrated in Table \ref{tab:session2}, all models achieved superior regression performance in decoding speaker-normalized pitch contours compared to raw pitch contours across the 6-speaker dataset, with the MAE metric showing an average improvement of 32.7\%. 
These findings collectively suggest that Mandarin perception relies on speaker-normalized relative pitch rather than absolute pitch, and critically, the normalization process occurs at the phoneme level during tonal perception.

\begin{table}[th]
  \caption{Results of different systems on session 1.}
  \label{tab:session1}
  \centering
  \begin{tabularx}{\columnwidth}{lXXXXXXXX} 
    \toprule
    \textbf{Systems} & \textbf{pitch} & \textbf{mae} & \textbf{rmae} & \textbf{gpe} \\
    \midrule
    VLAAI & raw & 22.6 & 16.5 & 31.1 \\
    VLAAI & norm & 22.5 & 16.1 & 29.7 \\
    HappyQuokka & raw & 22.3 & 16.3 & 30.6 \\
    HappyQuokka & norm & 20.8 & 14.8 & 24.1 \\
    Wavenet & raw & 18.6 & \textbf{13.2} & 21.9 \\
    Wavenet & norm & 18.5 & \textbf{13.0} & \textbf{21.4} \\
    \midrule
    Proposed & raw & \textbf{18.4} & \textbf{13.2} & \textbf{21.4} \\
    Proposed & norm & \textbf{18.4} & 13.1 & 22.0 \\
    \bottomrule
  \end{tabularx}
\end{table}

\begin{table}[th]
  \caption{Results of different systems on session 2.}
  \label{tab:session2}
  \centering
  \begin{tabularx}{\columnwidth}{lXXXXXXXX} 
    \toprule
    \textbf{Systems} & \textbf{pitch} & \textbf{mae} & \textbf{rmae} & \textbf{gpe} \\
    \midrule
    VLAAI & raw & 51.8 & 31.6 & 56.9 \\
    VLAAI & norm & 36.3 & 22.1 & 46.2 \\
    HappyQuokka & raw & 49.9 & 32.1 & 55.0 \\
    HappyQuokka & norm & 34.6 & 21.3 & 42.1 \\
    Wavenet & raw & \textbf{46.7} & \textbf{29.0} & \textbf{49.8} \\
    Wavenet & norm & 31.3 & \textbf{19.2} & 35.2 \\
    \midrule
    Proposed & raw & 48.9 & 31.1 & 53.1 \\
    Proposed & norm & \textbf{30.5} & 19.5 & \textbf{34.7} \\
    \bottomrule
  \end{tabularx}
\end{table}

\subsection{Ablation Study}
In this section, we conducted several ablation studies to analyze the impact of each module on the performance of the proposed model. The two main modules: the Convolutional Embedding and the Feature Encoder, which were either fully or partially removed to observe their individual effects. The resulting performance metrics on session 1 are presented in Table \ref{tab:Ablation}.

Compared with the proposed method, removing any module will bring an obvious performance degradation, implying the necessity of both module.
Removing Convolutional Embedding (w/o TCE/CCE/CE) results in a greater performance drop compared to removing Feature Encoder (w/o TFE/CFE/FE), indicating that Convolutional Embedding contributes more significantly to performance. 
However, it should be noted that Feature Encoder can further enhance the complementarity of the learned representations. 
For the Convolutional Embedding, a system without the Channel Branch (w/o CCE) achieves better results than one without the Temporal Branch (w/o TCE). 
This suggests that rich temporal local information is crucial for the outcome.

\begin{table}[th]
\tabcolsep=0.6cm
  \caption{Ablation results of the proposed model with raw pitch contours on session 1. Here, CE, TCE, and CCE denote Convolutional Embedding, its Temporal Branch, and Channel Branch, respectively. FE, TFE, and CFE represent Feature Encoder, Channel Transformer Encoder, and Temporal Transformer Encoder, respectively.}
  \label{tab:Ablation}
  \centering
  \begin{tabular}{{lcccc}} 
    \toprule
    \textbf{Systems} & \textbf{mae} & \textbf{rmae} & \textbf{gpe} \\
    \midrule
    Proposed & \textbf{18.4} & \textbf{13.2} & \textbf{21.4} \\
    \midrule
    w/o TCE & 20.7 & 14.5 & 23.9 \\
    w/o CCE & 19.0 & 13.8 & 22.2 \\
    w/o CE & 21.2 & 15.4 & 26.1 \\
    w/o TFE & 19.0 & 13.6 & 22.3 \\
    w/o CFE & 18.6 & 13.4 & 21.6 \\
    w/o FE & 19.9 & 14.1 & 22.7 \\    
    \bottomrule
  \end{tabular}
\end{table}
\vspace{-0.25cm}

\section{Conclusion}
\vspace{0.1cm}

In this paper, we demonstrate that the brain's perception of Mandarin pitch contours is independent of individual speakers' pitch ranges. We recorded a Mandarin auditory EEG dataset and proposed a novel auditory EEG decoding model, CE-ViViT. Results on session 1 and session 2 show that the proposed model achieves performance comparable to state-of-the-art methods. In the future, we will explore whether sentence-level context affects speaker-normalized pitch perception.



\bibliographystyle{IEEEtran}
\bibliography{mybib}

\end{document}